\documentclass[sigplan,screen,nonacm=true]{acmart}
\usepackage[T1]{fontenc}
\usepackage{algorithm}
\usepackage{algorithmic}

\usepackage{bm,calrsfs}

		\setcopyright{none}

\newcommand{\mE}{\mathbb{E}}
\newcommand{\mR}{\mathbb{R}}

\newcommand{\mQ}{\mathbb{Q}}
\begin{document}

\title{Stock Volume Forecasting with Advanced Information by Conditional Variational Auto-Encoder}

\author{Parley R Yang}
\email{ry266@cam.ac.uk}
\affiliation{%
  \institution{Faculty of Mathematics, University of Cambridge}
  \city{Cambridge}
  \country{UK}
}

\author{Alexander Y Shestopaloff}
\affiliation{%
	\institution{School of Mathematical Sciences, Queen Mary University of London}
	\city{London}
	\country{UK}
}

\begin{abstract}

We demonstrate the use of Conditional Variational Encoder (CVAE) to improve the forecasts of daily stock volume time series in both short and long term forecasting tasks, with the use of advanced information of input variables such as rebalancing dates. CVAE generates non-linear time series as out-of-sample forecasts, which have better accuracy and closer fit of correlation to the actual data, compared to traditional linear models. These generative forecasts can also be used for scenario generation, which aids interpretation. We further discuss correlations in non-stationary time series and other potential extensions from the CVAE forecasts.

\vfil

\end{abstract}

\maketitle

	\pagebreak
	
\section{Introduction}

\subsection{Motivation}\label{MLTF}
Let $Y_{t+k|t}$ be the forecast of variable $Y$ at time $t+k$, given information available up to time $t$. For instance, in the case of daily stock data where $Y_t$ indicates the end of day stock price of day $t$, $Y_{t+1|t}$ would be the forecasted stock price on day $t+1$ based on information up to time $t$. 

Advanced information concerns with the notion of information available up to time $t$. In linear time series, such information is mainly the past lags of the data (autoregressive terms) and error terms, together with other features observed up to time $t$. Advanced information acknowledges the future state of some variables at time $t$. For instance, there is an explicit rule of when Stoxx index rebalancing would occur \cite{Stoxx24} and this information may concern the future state $t+k$, but is known at time $t$. If we write $RB_t$ as the indicator of whether day $t$ is a rebalancing day for Stoxx index, then $RB_{t+k}$ is known at time $t$ for all $k$. It can therefore be helpful to utilise such advanced information at time $t$ to construct a forecast of $Y$ at time $t+k$, especially when $k$ is large. However, how to incorporate such an advanced information to the time series model becomes a crucial question, in light of the potential non-linear effect this information has on $Y_{t}$. In this paper, we use Conditional Variational Auto-Encoder (CVAE) to account for this in a non-linear modelling and forecasting setting.

The notion of long term forecasting concerns $k$ being large, for instance, the stock price two weeks (ten trading days) later, $Y_{t+10|t}$. In the analysis of linear stationary time series (e.g. ARMA and VARMA), the long term forecasts converge to the unconditional expectation due to the assumption of stationarity\footnote{Technically, the conditional forecast $Y_{t+k|t} = \mE[Y_{t+k}|{\mathcal F}_t] \to \mE[Y]$ as $k\to\infty$ where ${\mathcal F}_t$ is the set of information up to time $t$, and $\mE[Y_t]=\mE[Y] \forall t$ due to the stationarity assumption.}. However, in the case of non-stationary time series, such convergence is not guaranteed, and can be of interests. 

Empirically, in the case of stock price, it may be convenient to put $Y_{t+10|t}$ or even $Y_{t+100|t}$ to some trend-stationary expectation or equivalent, as the difference of stock prices (either $Y_t - Y_{t-1}$ or $\log(Y_t )- \log(Y_{t-1})$) are often modelled as a stationary time series. In the case of stock volume, however, it is not as convenient to assert such stationary expectation, as stock volumes are often affected by extraordinary shocks, e.g. company announcements and index rebalancing. It is therefore a harder task, both in modelling and forecasting, to provide analysis for stock volume time series. 

Furthermore, forecasting stock volume in a non-linear fashion helps the pricing of various financial derivatives, such as stock buyback contracts, which heavily depend on market volumes for periods covered by such contracts.

\subsection{Literature Review}

There are some recent literature on time series with the use of neural network, such as Time Series generative adversarial networks \cite{YJv19} and Generative Time Series with bi-directional Variational Auto-Encoder (VAE) \cite{LLWD22}. We take a similar approach in designing the latent space and building Gaussian assumptions conditional on latent space. However, we take into account the advanced information and utilise such information to improve the quality of our forecast, with the incorporation of latent variables. We also use CVAE to enable scenario generation and interpretation. To this end, we engage with technical papers on CVAE \cite{doersch2016tutorial, CVAE2015} for modelling and derivations under a time series setting.

In terms of the literature of advanced information, there are no direct reference of this terminology in time series, however, we find similar concepts in Bayesian Time Series \cite{Berliner96,tsay2018nonlinear}, where one updates the state equations based on some signals. Such signal may be learnt or stated in advanced for the model --- advanced information may be considered as a set of variables where we know the existence of its value in advance, and need not be learnt for the duration of present and future forecasting period.

The empirical need for stock volume forecasting can be motivated in the general financial machine learning literature \cite{de2018advances}, though most of the recent literature focus on the microstructure statistics, such as in the case of limit order book \cite{CKS13}. The role of stock volume time series in buyback contract pricing has been mentioned in recent literature \cite{GIJ20, HHP22}, which introduce practical forecasting problems that we aim to  empirically contribute to.

\subsection{Contributions}
Our overall contributions can be summarised into threefold. Firstly, we identify the class  of problem of forecasting with advanced information, which can be considered as an expectation computation based on a richer information set --- this  is practically modelled with non-linear interactions in a CVAE architecture of neural networks, which additionally enables generative forecasting. This is elaborated in section \ref{Method}.

Secondly, we demonstrate the capability for event-driven interpretations and alternative scenario generation. This utilises the generator aspect of CVAE to answer questions such as what happens on the special occasion and what are the alternative scenarios. By generating the forecast paths under different conditional values, we are able to answer these questions and henceforth provide interpretation to the model. These are provided in addition to the traditional model evaluation metrics such as Mean Squared Error (MSE) and correlation matrix, which are detailed in section \ref{Emp}.

Lastly, we contribute to the empirical literature on daily stock volume forecasting, which is studied across the incumbents of EURO STOXX 50 index, which is a cluster of 50 high market-capitalisation stocks listed in the Eurozone. This contributes to the demand of long term forecasting in empirical finance, such as for buyback contracts.

\subsection{Notations}
Most of the notations are explained when first introduced. Common annotations are as follows. $N(\cdot, \cdot)$ indicates the Gaussian Distribution, $I_q$ indicates a q-dimensional identity matrix, upper case letters usually denote objects being a random variable, whereas lower case letters usually denote data observed at specific values. $\mE[\cdot]$ denotes the expectation operator, $V(\cdot)$ denotes the variance operator, with $Cov(\cdot,\cdot)$ as the covariance operator, and $Corr(\cdot,\cdot)$ as the correlation operator. Conditional expectation $\mE_{X \sim P}[Q(X)]$ refers to the expectation of $Q(X)$ conditional on $X\sim P$.

\section{Methodology: From Non-Linear Modelling To Generative Algorithm For Forecasting}\label{Method}

\subsection{Time series in a CVAE context}

	We first give an overview of the modelling assumptions of CVAE, and put time series in such a context.

	Let $X, Y$ be random variables. $Y \in \mR^d$ and $X \in \mR^p$. Let $Z \in \mR^{q}$ be a latent variable with distribution $Z \sim N(0, I_{q})$. Let $D$ be the dataset consisting paired observations of $(X,Y)$. The key assumptions are: conditional distribution of the output given input and latent variable as Gaussian with non-linear and unknown mean, written as 	\begin{equation}\label{VAEmodel}
		Y|X,Z \sim  N(f(X,Z), \sigma^2 I_d)
	\end{equation}
	for some unknown function $f:  \mR^p \times \mR^q \to \mR^d $, and conditional distribution of latent variable given the observed data ($X,Y$) is Gaussian, written as \begin{equation}\label{VAEmodel2}
		Z| X,Y \sim N(\mu(X,Y), \Sigma(X,Y))
	\end{equation} 
	
	A CVAE is a tuple $(\hat{f}^{en}, \hat{f}^{de})$ where $\hat{f}^{de}:  \mR^p \times \mR^q \to \mR^d $ is a decoder which induces a probability distribution $\hat{P}(Y|X,Z)$ according to \autoref{VAEmodel} and an encoder is a function $\hat{f}^{en}:  \mR^p \times \mR^d \to \mR^q \times (0,\infty)^q $ which induces the moments of the Gaussian distribution $ Z| X,Y$ according to  \autoref{VAEmodel2}, with the assumption that $\Sigma(X,Y)$ can be written as a diagonal matrix with positive entries on all diagonals.

	Further to this, time series $y_t, x_t$ are seen as observation of random variables $Y, X$ at time $t$, and by deliberate modelling, we aim to train a CVAE $(\hat{f}^{en}, \hat{f}^{de})$ that enables the forecast of future states of $y_{t+k}$ conditional on $x_{t+1}$. For example, let $x_t = y_{t-1}$, then $y_{t+1|t} =\mE_{Z\sim N(0,I_q), X=y_t}[ N(\hat{f}^{de}(X,Z), \sigma^2 I_d)] = \mE_{Z\sim N(0,I_q)}[ N(\hat{f}^{de}(y_t,Z), \sigma^2 I_d) ] = \mE_{Z\sim N(0,I_q)}[\hat{f}^{de}(y_t,Z)]$. Iteratively for $k\geq 2$, we have $y_{t+k|t} =  \mE_{Z\sim N(0,I_q)}[\hat{f}^{de}(y_{t+k-1|t},Z)]$.
	
	The training of a CVAE is done by using two Neural Network architectures $F_1, F_2$ to define the functions ${f}^{en} \in F_1, {f}^{de}\in F_2$, followed by gradient methods and other classical optimisation techniques to maximise $\mE_{X, Y \sim D}[\log(P(Y|X))]$ through conditional marginalisation over $Z$. Derivations and other technical remarks are written in appendix \ref{techrem}.
	
\subsection{Generative scheme from CVAE}

For a given decoder $\hat{f}^{de}$ and conditional variable $x_t$, the distribution $ \mE_{Z\sim N(0,I_q), X=x_t}[ N(\hat{f}^{de}(X,Z), \sigma^2 I_d)]$ can be approximated by the generative scheme of $S$ samples 
\begin{align}
	\forall s \in [S], & \texttt{ draw } z_s \sim N(0,I_q), \nonumber \\
	 & \texttt{ then draw } y_t^s \sim N(\hat{f}^{de}(x_t,z_s), \sigma^2 I_d) \label{VAEgen}
\end{align} 

Now, through \autoref{VAEgen}, we may compute the approximated expectation by taking the average across samples  $\{y_t^s\}_{s \in [S]}$. 

Putting this scheme in a forecasting regime, we may generate forecasts $\{y_{t+1|t}^s\}_{s \in [S]}$ and thereafter $\{y_{t+k|t}^s\}_{s \in [S]}$ for a generalised $k$. For a given horizon of forecast $K$, we call a forecast path as the vector $y^s_{t+\cdot | t}:= (y_{t+1|t}^s, y_{t+2|t}^s, ...,y_{t+K|t}^s)$, and the average forecast path $\bar{y}_{t+\cdot | t}:= (\bar{y}_{t+1|t}, \bar{y}_{t+2|t}, ..., \bar{y}_{t+K|t})$ where each entry $\bar{y}_{t+k|t}$ is the arithmetic average over the generated samples $\{y_{t+k|t}^s\}_{s \in [S]}$.

\subsection{A special forecasting scenario: advanced information}\label{AdvInfo}

	In this section, we formaly introduce, and give examples to, the notion of advanced information.

	Given $K \geq 1$, we are interested to forecast random variable $Y_{t+k}, k \in [K]$ given information availble up to time $t$. Let $X_t = (X_t^0, X_t^1)$ where $X_{t+k+1}^0$ is always known at time $t$, for $k \in [K]$ but only up to $X_{t+1}^1$ is known at time $t$, not $X_{t+2}^1$ or anything further. We say $X^0$ is an advanced information and $X^1$ is an ordinary information. Intuitively, $X^1$ is the part of information where we only know up to the time we are requested to forecast, as was commonly assumed in classical linear time series models, whereas $X^0$ is the nuanced part where we know some information ahead of time. Then, we consider filtration ${\mathcal F}^*_t:={\mathcal F}^0_{t+K} \times {\mathcal F}^1_{t}$  where ${\mathcal F}^0_{t+K}$ is the filtration generated by $X_{t+K}^0$ and $ {\mathcal F}^1_{t}$ is the filtration  generated by $X_{t+1}^1$. Forecasting with advanced information concerns investigating the conditional distributions $\mQ^{*k}(\cdot | {\mathcal F}^*_{t})$ where $k \in [K]$ and $\mQ^{*k}(S | {\mathcal F}^*_{t})$ measures the probability of $Y_{t+k} \in S$ given ${\mathcal F}^*_{t}$. The expected forecast takes the form of $Y_{t+k|t}:=\int Y_{t+k} d\mQ^{*k}(Y_{t+k} | {\mathcal F}^*_{t})$ .
	
	To give some contextual remark, the inclusion of $X^0$ can often be considered as properties of the time series, such as category of variables. Say $Y \in \mR$ and some of the data belong to group 1 whereas the other belong to group 2, then $X^0 \in \{0,1\}^2$ can be a one-hot indicator function which outputs $(1,0)$ for group 1 and $(0,1)$ for group 2. In this case, when forecasting, $X^0$ is always known ahead of time (with effectively infinitely large $K$). This means many panel data models fall into the scenario of advanced information, as the category of the response can be considered as a known property, henceforth advanced information. Likewise for seasonality, where such category could be considered as an information exploitable ahead of the forecasting time.
	
	Another example is rebalancing, as was motivated in section \ref{MLTF}, that rebalancing dates are known ahead of the desired forecasting time, so $X^0$ may be an indicator of rebalancing date which outputs $1$ if the date is a rebalancing date, and $0$ otherwise.
	
	As a remark, $X^0$  may also be a special case for cointegration, as a source of drift. This may fall into the wider literature of 'common trend removal' in cointegration analysis.
	
	\subsection{Forecasting algorithms using advanced information and CVAE}
		\begin{algorithm}
		\caption{Iterative Forecasting with Advanced Information (General)}
		\label{alg:IFA}
	\begin{flushleft}
		\textbf{Input}: $t$ (time the forecast is requested), $S$ (number of samples), $\hat{f}^{de}$ (a trained decoder), $K$ (desired forecasting horizon), $\{x_\tau^0\}_{\tau = t+1}^{t+K}$ (advanced information), $x_{t+1}^1$ (ordinary information). 		\\ 
	\textbf{Output}: Simulated forecast paths $\{(k, y_{t+k|t}^s): k \in [K] \}_{s \in [S]}$		
	\end{flushleft}
		\begin{algorithmic}[1]
			\STATE Draw S samples from the distribution $ \mE_{Z\sim N(0,I_q), X^0=x^0_{t+1}, X^1=x^1_{t+1}}[ N(\hat{f}^{de}(X^0, X^1, Z), \sigma^2 I_d)]$ according to \autoref{VAEgen}. These samples are denoted  as $y_{t+1|t}^s$ for $s \in [S]$.
			\FOR{$\tau \in \{2,3,..., K\}$}
			\STATE Update $x^1_{t+\tau|t}$ with $\{y_{t+\tau-1|t}^s : s\in [S]\}$
			\STATE Draw S samples from the distribution $ \mE_{Z\sim N(0,I_q), X^0=x^0_{t+\tau}, X^1=x^1_{t+\tau |t}}[ N(\hat{f}^{de}(X^0, X^1, Z), \sigma^2 I_d)]$ according to \autoref{VAEgen}. These samples are denoted  as $y_{t+\tau|t}^s$ for $s \in [S]$.
			\ENDFOR
			
			\RETURN $\{(k, y_{t+k|t}^s): k \in [K] \}_{s \in [S]}$
		\end{algorithmic}
	\end{algorithm}
	
	In algorithm \ref{alg:IFA}, we present the iterative forecasting algorithm with advanced information, under the setting where decoder ${f}^{de}$ takes $X^0, X^1, Z$ as input. The algorithm uses \autoref{VAEgen} for the CVAE generation and serves as a practical algorithm for forecasting with advanced information. 
	
	Another algorithm as a special case for AR(1)-type of forecasting (the case where $X^1_{t} = Y_{t-1}$) is presented in algorithm \ref{alg:IFA2}, where there are more specified approach in handling $X^1$, as would be done in classical linear time series models. This is also the exact algorithm used in the empirical applications section next.
	
		\begin{algorithm}
		\caption{Iterative Forecasting with Advanced Information and 1-lag Autoregressive Ordinary Information}
		\label{alg:IFA2}	\begin{flushleft}
			\textbf{Input}:  $t$, $S$, $\hat{f}^{de}$, $K$, $\{x_\tau^0\}_{\tau = t+1}^{t+K}$, $x_{t+1}^1$, and $y_t$ (last observation of $Y$ at the time of forecast)	\\ 
			\textbf{Output}:  $\{(k, y_{t+k|t}^s): k \in [K] \}_{s \in [S]}$		
		\end{flushleft}
		\begin{algorithmic}[1]
			\STATE Draw S samples from the distribution $ \mE_{Z\sim N(0,I_q), X^0=x^0_{t+1}, X^1=y_t}[ N(\hat{f}^{de}(X^0, X^1, Z), \sigma^2 I_d)]$ according to \autoref{VAEgen}. These samples are denoted  as $y_{t+1|t}^s$ for $s \in [S]$.
			\FOR{$\tau \in \{2,3,..., K\}$}
			\STATE Average $\hat{y}_{t+\tau-1|t} = \frac{\sum_{s\in [S]}y_{t+\tau-1|t}^s}{S}$
			\STATE Draw S samples from the distribution $ \mE_{Z\sim N(0,I_q), X^0=x^0_{t+\tau}, X^1=\hat{y}_{t+\tau-1|t} }[ N(\hat{f}^{de}(X^0, X^1, Z), \sigma^2 I_d)]$ according to \autoref{VAEgen}. These samples are denoted  as $y_{t+\tau|t}^s$ for $s \in [S]$.
			\ENDFOR
			
			\RETURN $\{(k, y_{t+k|t}^s): k \in [K] \}_{s \in [S]}$
		\end{algorithmic}
	\end{algorithm}
	
\section{Empirical Application: Daily Stock Volume Forecasting}\label{Emp}

In this section, we model and forecast daily stock volume for 50 European stocks which were components of Euro Stoxx 50 as of the end of year 2023.

\subsection{Data Availability and Processing}
Daily stock volume data were obtained from Yahoo Finance. We split the training and testing as start of year 2021 to end of year 2022, and start of year 2023 to end of June 2023, respectively. We use the traning data to normalise the time series --- that is, for each stock, we find mean and variance in training period, then de-mean and unify the variance\footnote{The $Y\mapsto \frac{Y-\hat{\mu}}{\hat{\sigma}}$ mapping, where $\hat{\mu}$ and $\hat{\sigma}$ are, respectively, the mean and variance estimated from their training period}, as one of the standard Machine Learning data processing procedure.

\begin{figure}[h]
	\centering
	\includegraphics[width=1\linewidth]{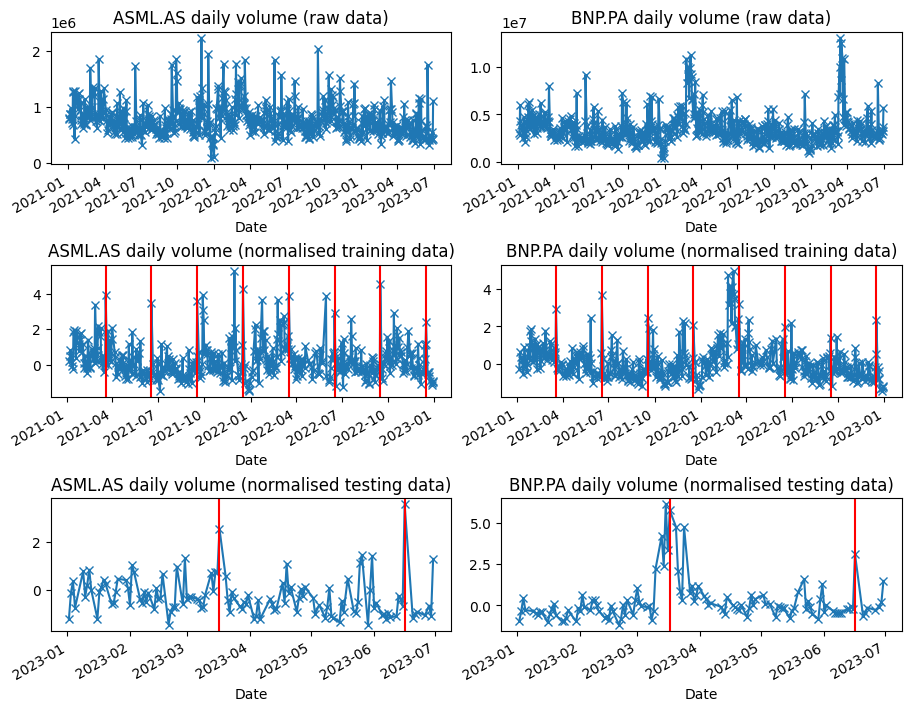}
	\caption{Illustrations of raw data and processed data for training and testing}
	\label{fig:tsplot}
\end{figure}

In addition to the volume data, we obtain stoxx rebalancing dates according to the STOXX index guide, and individual stock categories (Location and Sector) from the index-tracking ETF, EUE. These are seen as advanced information in modelling and foreacsting, as the information attributed are either known ahead of time, or believed to be unchanged throughout the forecasting horizon.

As an illustration on some of these data, in \autoref{fig:tsplot}, we plot two of the stocks (ASML.AS and BNP.PA) in both row data and processed data. The processed data rows (the second row for training and the third row for testing) also contains vertical red lines highlighting the rebalancing dates. It is visually clear that rebalancing dates do tend to co-occur with a higher-than-average volume --- this is something we would wish our model to capture, both in in-sample modelling and out-of-sample forecasting.

We annotate the scalar $y(i)_t$ as the normalised observed stock volume in day $t$ and for stock $i \in [N]$ where $N=50$, and the vector $\bm{y}_t \in \mR^{50}$ as the vector of observed stock volume in day $t$ across all stocks. Where there are missing observations in a day, we drop such an observation. 

As for features, $RB_t\in \{0,1\}^3$ denotes a three-dimensional one-hot encoder indicating the day relative to rebalancing day --- in particular, $RB_t = (1,0,0)$ indicates the last observation before rebalacing day, $RB_t = (0,1,0)$ indicates the rebalacing day, and $RB_t = (0,0,1)$  indicates the next observation after rebalacing day. $RB_t = (0,0,0)$ for all other days. The feature $DoW_t\in \{0,1\}^5$ denotes the day of the week of the observation. Another set of features are $Sector(i)_t \in \{0,1\}^{10}$ and $Location(i)_t \in \{0,1\}^7$ which are one-hot encoders for the sector of the stock $i$ and location of stock $i$ respectively. They are invariant over time and henceforth considered as advanced information.

\subsection{Forecasting Tasks}
We would like to accomplish two forecasting tasks through our models. 

The long term forecasting concerns the scenario where we are given large $K$ and a fixed $t$ (end of the training sample), and we request to forecast $y(i)_{t+k|t}$ for $k \in [K], i \in [N]$. In this dataset, we have $K=120$ representing the first half of 2023 (business days and after removal of days with empty observations due to holidays or data source error).

The short term rolling forecast concerns the scenario of week-long forecasts on multiple periods, hence $K\leq 5$ ($K=5$ is often the case where the forecast run from Monday to Friday --- in some cases such as the public holiday, it would reduce to $K<5$). Given a set of time points $t_1,...,t_u$ indicating end of the current week, and some $k_1,..., k_u \leq 5$ which indicates the duration of forecasts from the start of the upcoming week to the end of the upcoming week, we  request to forecast $y(i)_{t+k^\prime|t}$ for $k^\prime \in [k], (t, k) \in \{(t_1,k_1),...,(t_u, k_u)\}, i \in [N]$.

\subsection{Summary of Models}
Recall the setting of CVAE as per \autoref{VAEmodel} and \autoref{VAEmodel2}. We model a univariate CVAE (U-CVAE) on each individual $y(i)_t$ and a multivariate CVAE (M-CVAE) on the vector $\bm{y}_t$.  W keep $q=1$ for simplicity.

For the U-CVAE, we have $Y(i)_t \in \mR$ and model \begin{align}\label{VAEmodel_CVAE}
	Y(i)_t |X(i)_t ,Z \sim &  N(f(X(i)_t,Z), \sigma^2), \\ &X(i)_t=(X(i)^0_t, X(i)^1_t), \ \ Z \sim N(0,1)
\end{align}
with $X(i)^0_t = (Sector(i)_t, Location(i)_t,DoW_t, RB_t)$ and  $X(i)^1_t = Y_{t-1}(i)$.

For the M-CVAE, we have $ \bm{Y}_t \in \mR^{50}$ and model \begin{align}\label{VAEmodel_MVAE}
	 \bm{Y}_t | X_t ,Z \sim &  N(f(X_t,Z), \sigma^2 I_{50}), \\ &X_t=(X_t^0, X_t^1), \ \ Z \sim N(0,1)
\end{align}
with $X_t^0 = (DoW_t, RB_t)$ and  $X_t^1 = \bm{Y}_{t-1}$.

For comparison, we also provide two baseline models: for the univariate baseline, we model and forecast $y(i)_t$ using ARMA(1,1); and for the multivariate baseline, we do so on $\bm{y}_t$ using VAR(1). \footnote{VARMA(1,1) faces insufficient data due to the number of parameters almost exceeding the amount of data available.}
\subsection{Evaluation Metrics}
\begin{figure}
	\centering
	\includegraphics[width=1\linewidth]{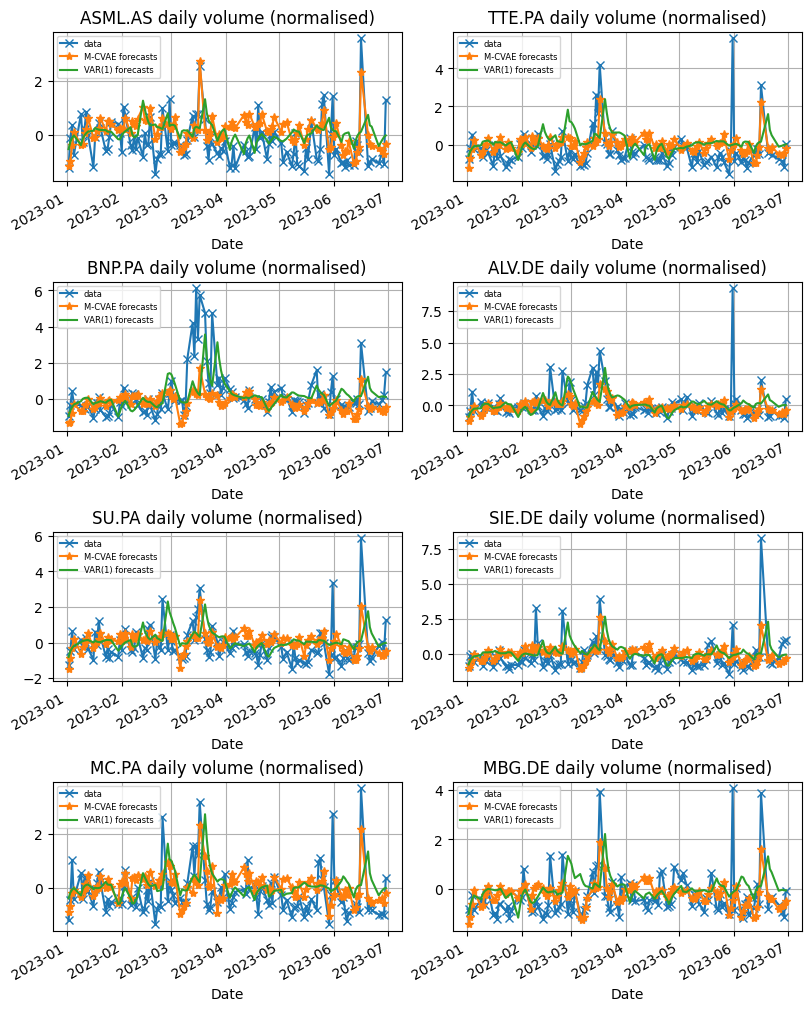}
	\caption{Short Term Rolling Forecasts: M-CVAE and VAR(1) Illustrations}
	\label{fig:rftsplot}
\end{figure}

\begin{figure}
	\centering
	\includegraphics[width=1\linewidth]{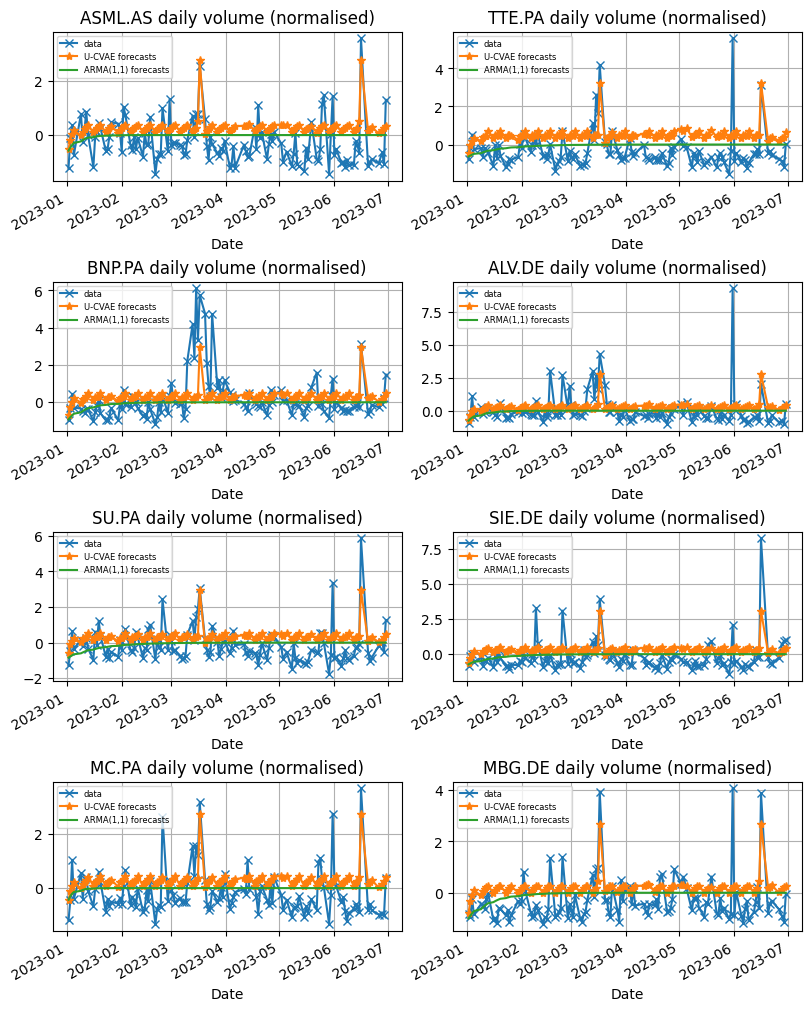}
	\caption{Long Term Forecasts: U-CVAE and ARMA(1,1) Illustrations}
	\label{fig:lftsplot}
\end{figure}

To evaluate the forecasts, we employ two core concepts in time series: Mean Squared Errors (MSE) and correlations. The computation of MSE is straightforward: given forecasts $\hat{y}(i)_{t+k|t}$ generated by the models, we compute the average MSE by comparing against the actual observation in the testing period: $y_{t+k}$, namely $(\hat{y}(i)_{t+k|t}-y_{t+k})^2$ averaged across the testing period (the precise notation varies depending on whether it is summed over short-time rolling forecast and long term forecast) for each stock $i$, denoted $MSE(i)$.  Summary of statistics (mean and median) can then be obtained for $MSE(i)$.

The correlation matrix may also be produced for $\hat{\rho}_{i,j} = corr(\hat{y}(i)_{t+k|t}, \hat{y}(j)_{t+k|t})$ against the correlation of the data ${\rho_{i,j}} =corr(y(i)_{t+k},y(j)_{t+k})$ (exact formulations are drawn in section \ref{corr_diss} where we further discuss the correlation of non-stationary time series). To summarise the difference, we apply average absolute value to obtain the difference, denoted CD for correlation difference \begin{equation*}
	CD(i) = \frac{1}{50} \times \sum_{j \in [50]} | \hat{\rho}_{i,j}  - \rho_{i,j}|
\end{equation*}
Summary of statistics (mean and median) can then be obtained for $CD(i)$.

The cross-correlation matrix is also considered, as one may wish to observe the difference between lagged correlation amongst the forecasts. In particular, we produce $\hat{\rho}^*_{i,j} = corr(\hat{y}(i)_{t+k|t}, \hat{y}(j)_{t+k+1|t})$ and compare against the one by data ${\rho^*_{i,j}} =corr(y(i)_{t+k},y(j)_{t+k+1})$. We then summarise the difference in the same fashion as was done in $CD(i)$. We denote this statistic as $CCD(i)$ for cross correlation difference.

\subsection{Summary of Performance}\label{SoP}
As a summary of the performance in forecasting, we provide evaluation metrics in \autoref{SummarytabLF} and \autoref{SummarytabRF} for the long term forecasting task and short term rolling forecasts respectively. Illustrations of these forecasts are plotted in \autoref{fig:rftsplot}  and \autoref{fig:lftsplot}, where we can see immediate difference between CVAE forecasts and their baseline models.

\begin{table}[h]
	\begin{tabular}{l|rrrr}
		\toprule
	&	U-CVAE & M-CVAE & ARMA(1,1) & VAR(1) \\
		\midrule
mean MSE & \textbf{0.887} & 0.888 & 0.923 & 0.981 \\
median MSE & \textbf{0.876} & 0.884 & 0.922 & 1.001 \\
mean CD & 0.466 & \textbf{0.391} & 0.441 & 0.492 \\
median CD & 0.435 & \textbf{0.374} & 0.421 & 0.458 \\
mean CCD &  \textbf{0.106} & 0.146 & 0.755 & 3.669 \\
median CCD &  \textbf{0.096} & 0.136 & 0.837 & 3.644 \\
\bottomrule
	\end{tabular}
	\caption{Performance of Long Term Forecasts}
		\label{SummarytabLF}
\end{table}

\begin{table}
	\begin{tabular}{l|rrrr}
		\toprule
		&	U-CVAE & M-CVAE & ARMA(1,1) & VAR(1) \\
		\midrule
mean MSE & 0.793 & \textbf{0.788} & 0.971 & 1.070 \\
median MSE &  \textbf{0.737} &0.761 & 0.989 & 1.118 \\
mean CD &  0.240 & 0.275 & \textbf{0.093} &0.193 \\
median CD & 0.227 & 0.258 &  \textbf{0.083} & 0.179 \\
mean CCD & \textbf{0.124} & 0.262 & 0.377 & 0.420 \\
median CCD & \textbf{0.105} & 0.271 & 0.392 & 0.421 \\
		\bottomrule
	\end{tabular}
	\caption{Performance of Short Term Rolling Forecasts}
	\label{SummarytabRF}
\end{table}

As a summary, we see that the CVAE forecasts do better job in both long term and rolling short term forecasting tasks, with the out-performance in MSE --- more significantly so in the short term forecasts, and good fit of CCD in all cases. Correlation matrix fit well, though under-performs their baseline counterparts in short term forecasting tasks. Illustrations of correlation matrices are plotted in \autoref{fig:RF_corr} in the appendix.

From the illustrations, we can further appreciate such out-performance in two folds: quite significantly so in long term forecasts, the CVAE takes into account the advanced information in both modelling and forecasts, and are able to project spikes in a non-linear time series fashion that matches some of the spikes in the actual observation. Additionally, in short term forecasts, baseline models tend to have heavy reliance on their lagged dependent variables, which creates problems as to over- or under-forecasts in their forecasting horizon, these are partly mitigated in CVAE forecasts as they moderate these with the trained parameters of advanced information and other features.

There are further potential improvements in short-term forecasts: linear baselines do well in the correlation fitting --- this may be seen in the lower part of \autoref{fig:RF_corr} and the CD entries in \autoref{SummarytabRF}. Despite having low accuracies, linear models still preserve the correlation structures in their forecasts, resulting in a better fit --- whereas the CVAE models, despite better fit, tend to have more variabilities in their forecasts, and consequentially tend to over-fit in correlation when comparing to the actual data. In cross correlation, however, CVAE models still significantly outperform linear baselines.

We further some of these interpretations by zooming into the forecasts and discuss alternative scenarios of the feature values in the next section.

\subsection{Decoder as Generator: feature interpretation and scenario generation}\label{DaG}

\begin{figure}
	\centering
	\includegraphics[width=1\linewidth]{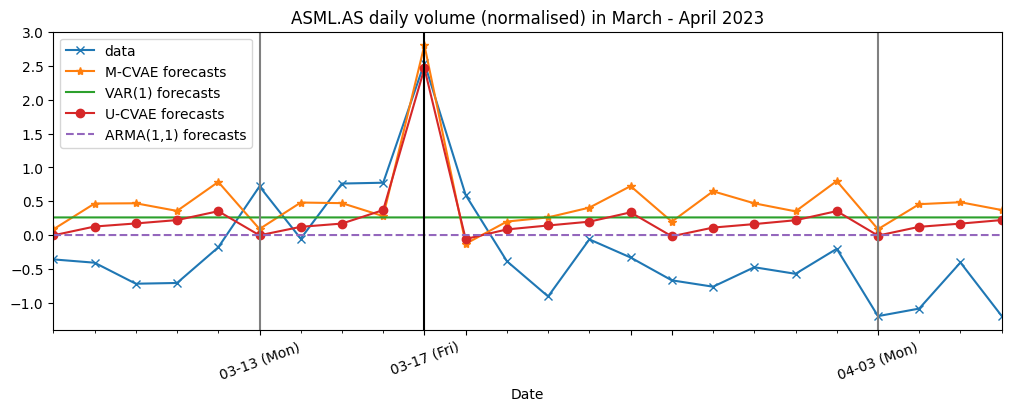}
	\caption{Long Term Forecasts: A zoomed-in plot for all models in March - April 2023, for ticker ASML.AS}
	\label{fig:ASML}
\end{figure}

\begin{figure}[h]
	\centering
	\includegraphics[width=1\linewidth]{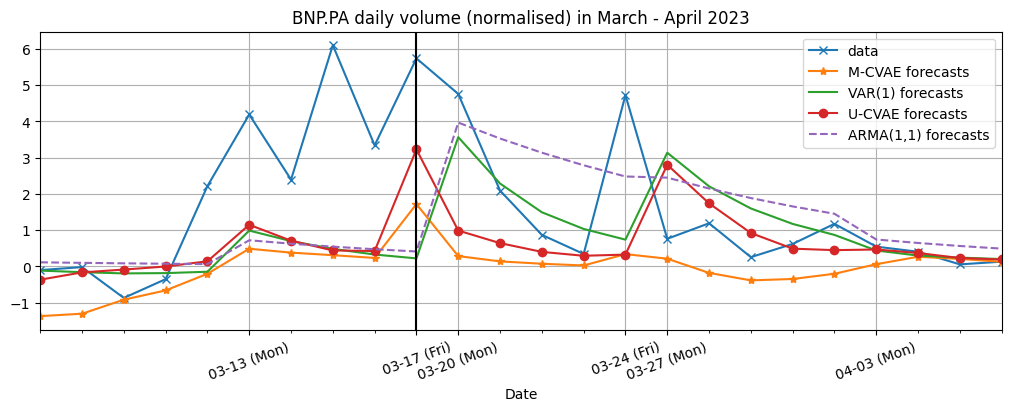}
	\caption{Short Term Rolling Forecasts: A zoomed-in plot for all models in March - April 2023, for ticker BNP.PA}
	\label{fig:BNP}
\end{figure}

In this section, we address two questions which help to appreciate the value of CVAE forecasts: How does RB affect the forecasts? And what's the  effect of the lagged dependent variable (similar to the IRF analysis in linear time series)? To engage with empirical data, we zoom into the long term forecast of ASML in March - April 2023 (illustrated in \autoref{fig:ASML}) to answer the first question, and the short term rolling forecast of BNP in the same period (\autoref{fig:BNP}) for the second question.

From \autoref{fig:ASML}, we can closely observe the ability of CVAE-generated forecasts to match the spike on the 17th March 2023, which is benefited from the advanced information of rebalancing date indicators (RB). Their corresponding baselines stay flat as the convergence of stationary forecasts would yield, when forecast period becomes large (the last observation being the end of year 2022).

Further to this, we may analyse the counterfactual of CVAE-generated forecasts in the case where there would be no rebalancing events in that week. To do this, we may augment the advanced information $X^0$ such that the $RB_t = (0,0,0)$ for the period of forecasts. We then generate the new forecast paths using algorithm \autoref{alg:IFA2} with the augmented advanced information $X^0$.
\begin{figure}[h]
	\centering
	\includegraphics[width=1\linewidth]{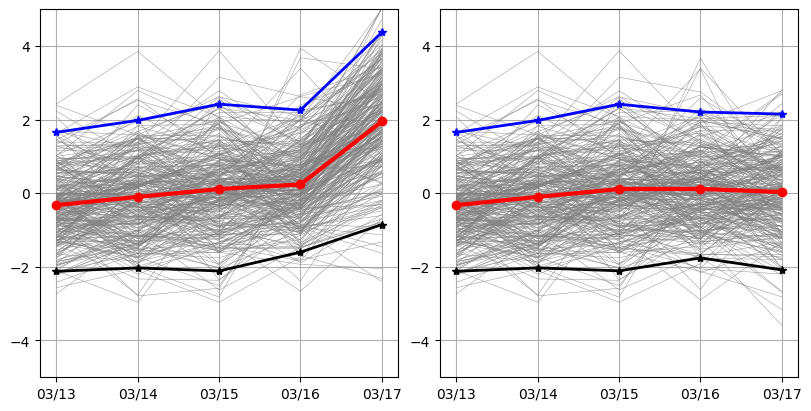}
	\caption{RB interpretation: ASML illustration (generated by U-CVAE)}
	\label{fig:rb}
\end{figure}

In \autoref{fig:rb}, we plot the paths (in grey) and their mean (in red), 97.5\% quantile (in blue) and 2.5\% quantile (in black). On the left panel are the forecasts with the actual $X^0$ where rebalancing happens on Friday, whereas on the right are the newly generated forecasts under the counterfactual with absence of rebalancing. There are slight difference on Thursday (16th March) paths as the one with rebalancing moves up and slightly tighter on the day before rebalancing date, and a huge difference on Friday (17th March) where rebalancing kicks up the forecast on the left panel --- the counterfactual in the absence of rebalancing date shows a relatively normal mean and upper and lower quantiles.

Using a similar convention, we may observe from \autoref{fig:BNP} that a spike in the data was observed on the 24th March. This comes in the episode of a higher-than-usual volume in March 2023 (mainly due to the European banking sector distress during that period). We zoom in to the last week of March (week commencing 27th March) for the analysis of lagged impact. As it is rolling forecasts, we can see U-CVAE and its baseline all picked up the last observed spike in their upcoming short term forecasts of 5 days. The spike on the 24th March was recorded at just below 5 --- it is tempting to seek for the impulse of this extraordinary observation by comparing against a counterfactual of the observation being at zero or at the negative of what was observed (just above -5). 

\begin{figure}
	\centering
	\includegraphics[width=1\linewidth]{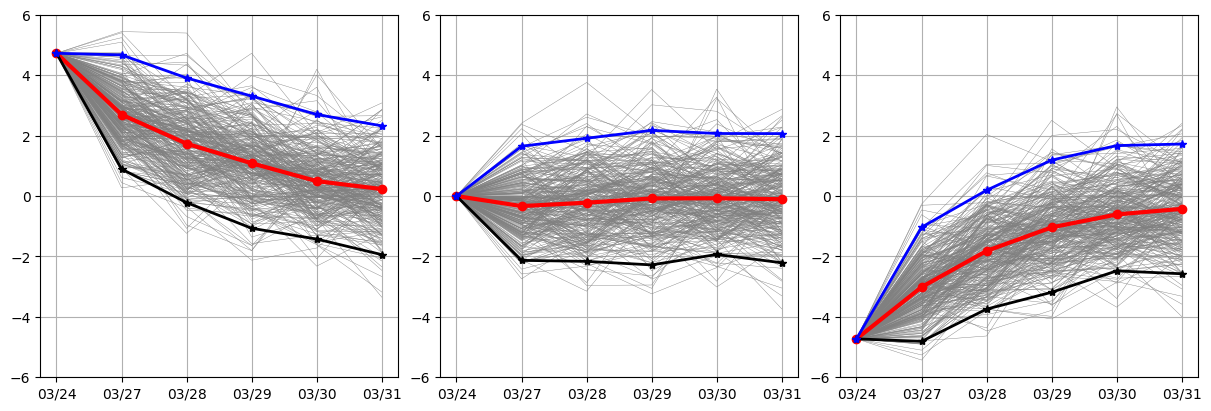}
	\caption{Lagged volume feature interpretation: BNP illustration (generated by U-CVAE)}
	\label{fig:li}
\end{figure}

To do this, two alternative paths are generated by replacing the $X^1_t=y_t$ part of  algorithm \autoref{alg:IFA2} to our desired value of  the counterfactual states. We plot such paths in \autoref{fig:li} with the augmented $y_t$ values. As visualised, the U-CVAE generated forecasts have a similar shape as are in the stationary time series analysis, where the paths converge back to the longer-term stationarity (around 0) when facing a upward or downward shock --- in reality, the upward shock was in effect, under which U-CVAE responded well, similar to how the baseline performed.

\section{Further Discussions}\label{Discuss}

\subsection{Path correlation in non-stationary time series}\label{corr_diss}
Recall that for a given k, a forecast path can be written as $y_{t+\cdot|t} := (y_{t+1|t},..., y_{t+k|t})$  and likewise for $y(i)_{t+\cdot|t}$, which takes the $i$-th entry, namely $y(i)_{t+\cdot|t} := (y(i)_{t+1|t},..., y(i)_{t+k|t})$. The correlation statistics takes the form of $$\rho_{i,j} = \frac{Cov(y(i)_{t+\cdot|t}, y(j)_{t+\cdot|t})}{\sqrt{V(y(i)_{t+\cdot|t}) V(y(j)_{t+\cdot|t})  } }$$ between paths $y(i)_{t+\cdot|t}$ and $y(j)_{t+\cdot|t}$. Now, to obtain this statistics, we consider the 
correlation of the average paths (CAP), defined over the average path $\bar{y}_{t+\cdot|t}$ generated. This can be expressed as 		\begin{equation}
\hat{\rho}^{CAP}_{i,j} =  \frac{Cov( \bar{y}(i)_{t+\cdot|t}, \bar{y}(j)_{t+\cdot|t}) }{\sqrt{V(  \bar{y}(i)_{t+\cdot|t}) V(  \bar{y}(j)_{t+\cdot|t})  } }
\end{equation}
Throughout this paper so far, we used CAP to report the correlation statistics. However, it can also be of interests to report another statistics, which concerns per-path correlation. To this end, we define the average correlation of paths (ACP), denoted $\hat{\rho}^{ACP}_{i,j} $, as below:
	\begin{equation}
\forall s, \hat{\rho}^s_{i,j} :=  \frac{Cov(y(i)^s_{t+\cdot|t}, y(j)^s_{t+\cdot|t})}{\sqrt{V(y(i)^s_{t+\cdot|t}) V(y(j)^s_{t+\cdot|t})  } }; \ \  \hat{\rho}^{ACP}_{i,j} = \frac{\sum_{s\in S} 	\hat{\rho}^s_{i,j} }{
	|S|}
\end{equation}

\begin{figure}[h]
	\centering
	\includegraphics[width=1\linewidth]{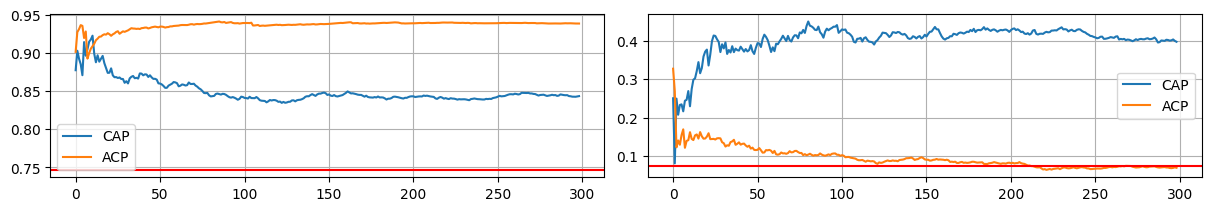}
	\caption{Correlation estimates (left) and cross correlation (right), with red line indicating the true data estimation}
	\label{fig:corr3}
\end{figure}

To empirically showcase the difference, we compute such statistics on the correlation and cross correlation between ASML and BNP, and plot the rolling estimations (expanding window of samples) on \autoref{fig:corr3} for short term rolling forecasts and under $\sigma=1$ configuration. Clearly, ACP and CAP converge to different values, with CAP having less bias approximating the true correlation and ACP having a better try on approximating the true cross correlation. A generalised set of cross correlation matrices are plotted in \autoref{fig:corr2} in the appendix, where ACP tends to fit well with the actual data while CAP mostly over-estimates the values.

It is in general not easy to derive the limiting distribution of the correlation variables when the sample size goes to infinity. Fixing $k$, under linear time series models (stationary ARMA or VAR in particular), each sample would be drawn from a Gaussian distribution, hence both covariance and variance would converge to $\chi^2$ in distribution. However, non-stationary time series does not have such convergence, making it hard to analytically showcase the limiting distribution.

Likewise, it may be tempting to discuss the limit when $k$ goes to infinity. Under linear time series, the forecast will be gradually concentrating to a point as $y_{t+k|t} \to \mE[Y] \text{ as } k \to \infty$ under the stationarity assumption, which leads to the unbiasedness of correlation estimate when $k \to \infty$, should all variables be stationary. However, non-stationary variables do not benefit from this, and ACP is conjectured to be different from CAP under the non-linear setting. One may derive the limiting distribution under simple regime-switching for this case.

\subsection{Other Extensions}
There are many other aspects of the CVAE forecasting schemes which are not  discussed extensively in this paper due to the length restriction. First and foremost, we have alternative generative schemes that worth consideration and demonstration --- instead of drawing from $$ \mE_{Z\sim N(0,I_q), X=x_t}[ N(\hat{f}^{de}(X,Z), \sigma^2 I_d)]$$we may wish to draw $Z$ from conditional distribution based on the previous observation, that is, from $$ \mE_{(Z\sim Z|X=x_{t-1}, Y=y_{t-1}), \ X=x_t}[ N(\hat{f}^{de}(X,Z), \sigma^2 I_d)]$$ This makes it possible to interact with the encoder, as the distribution of $Z|X,Y$ is learnt by the encoder function. In fact, in some literature, this is the main generating method \cite{LLWD22}. 

Additionally, as was introduced in section \ref{AdvInfo}, advanced information is a more generalised concept of forecast measures, instead of just point forecasts --- hence one could look into interval forecasting, in addition to our investigation of mean forecast. This will then be comparable to traditional linear models. An attempt is made in section \ref{DaG} when plotting the upper and lower quantiles of the simulated paths, though more thorough extension would be desired.

There are many other aspects of machine learning techniques on the neural network architectures that could be extended, including the increase of latent dimension and alternative architectures such as convolutional neural networks. These could act as better forms of approximation to the true non-linear and non-stationary nature of the data.

\section{Conclusion}
In this paper, we first identify the class of problem of time series forecasting with advanced information, which can be related to many problems in time series and finance, including stationarity, panel data, and stock volume forecasting. A CVAE architecture is introduced to model such time series,   We further investigate the case for daily stock volume forecasting, and found the CVAE generated forecasts more competitive than traditional linear models. The CVAE forecasts may be further generated for different scenarios of input features, creating a possibility to interpret feature inputs and generate various scenarios. Various extensions are then discussed in light of potential challenges which can be encountered in non-linear time series.

\vfill
\pagebreak

\appendix
\begin{figure*}[h]
	\centering
	\includegraphics[width=\linewidth]{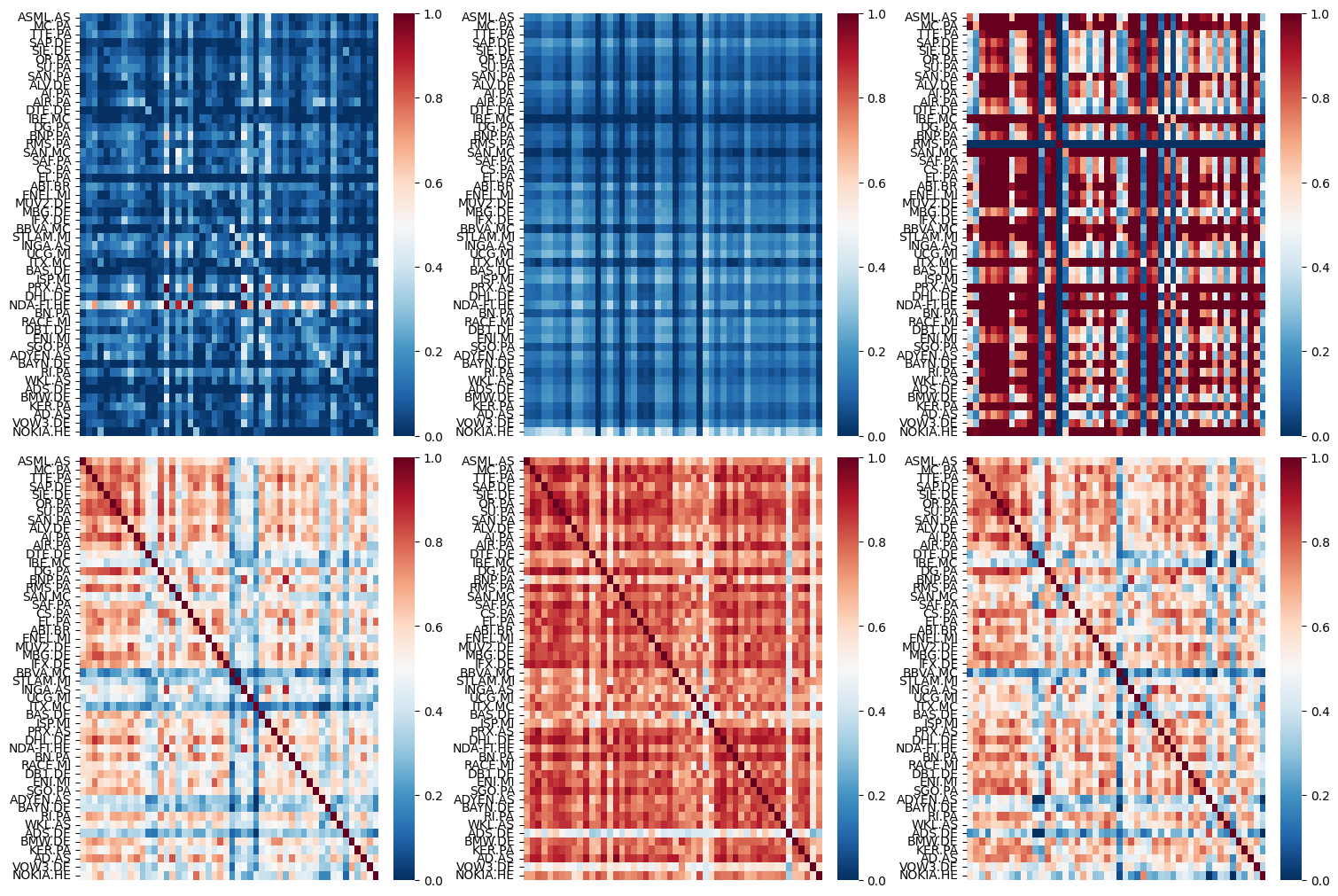}
	\caption{Cross Correlation Matrix of Long Term Forecasts (Upper) and Correlation Matrix of Short Term Rolling Forecasts (Lower): Data (left), U-CVAE (centre), and ARMA(1,1) (right)}
	\label{fig:RF_corr}
\end{figure*}

\section{Appendix}

\subsection{Technical remarks on the traninig of CVAE}\label{techrem}

\subsubsection{ERM for training}

The empirical risk minimisation for training the CVAE goes as follows. We aim to minimise $\mE_{X, Y \sim D}[\log(P(Y|X))]$ based on the assumptions in \autoref{VAEmodel} and \autoref{VAEmodel2}. Let $P(Y|X,Z)$ denote the probability distribution as specified in \autoref{VAEmodel}, let $P(Z) = N(0,I_q)$ be the distribution of $Z$, let $Q(Z|X,Y)$ be the distribution of $Z$ conditional on $X,Y$ as specified by \autoref{VAEmodel2}, and let $P(Z|X,Y)$ be the conditional distribution obtained by Bayes rule, i.e. $P(Z|X,Y) = \frac{P(Y|X,Z) P(Z|X)}{P(Y|X)}$

Write $KL(\cdot || \cdot)$ as the KL divergence, then observe, by Bayes' Rule
\begin{align*}
&KL(Q(Z|X,Y) || P(Z|X,Y)) \\
= &\mE_{Z\sim Q(Z|X,Y)}[\log(Q(Z|X,Y)) - \log(P(Z|X))\\&  + \log(P(Y|X)) - \log(P(Y|X,Z)) ] \\
= & KL(Q(Z|X,Y) || P(Z|X)) \\&  + \log(P(Y|X)) - \mE_{Z\sim Q(Z|X,Y)}[\log(P(Y|X,Z))]
\end{align*}
Rearranging get
\begin{align*}
\log(P(Y|X)) =	& \mE_{Z\sim Q(Z|X,Y)}[\log(P(Y|X,Z))] \\& + KL(Q(Z|X,Y) || P(Z|X,Y)) \\&- KL(Q(Z|X,Y) || P(Z|X))  \\
	\geq  & \mE_{Z\sim Q(Z|X,Y)}[\log(P(Y|X,Z))]  \\& - KL(Q(Z|X,Y) || P(Z|X)) 
\end{align*}
The last line is also known as Variational Lower Bound. Instead of interacting with $\log(P(Y|X)) $, we interact with the Variational Lower Bound. Same as the literature, we assume $P(Z|X)=P(Z)$ for latent variable to be independent of the input \cite{CVAE2015}. This enables the KL divergence term to be written in explicit form, which is 
\begin{align*}
&KL(Q(Z|X,Y) || P(Z)) \\
= & \frac{||\mu(X,Y)||^2_2+tr(\Sigma(X,Y)) - \log(\det(\Sigma(X,Y))) }{2} 
\end{align*}

Now, given dataset $D=\{(x_i, y_i)\}_{i \in I}$, we minimise the empirical version of the bound, namely 
$$\mE_{X, Y \sim D}[ \mE_{Z\sim Q(Z|X,Y)}[\log(P(Y|X,Z))] - KL(Q(Z|X,Y) || P(Z))] $$

The first term can be further approximated with simulated samples $z^i_1,...,z^i_S \sim Q(Z|x_i,y_i) \forall i$ \begin{equation*}
\sum_{i\in I} \sum_{s \in S} \frac{\log(P(y_i|x_i,z^i_s))}{|I| |S|}
\end{equation*}
And by \autoref{VAEmodel}, the probability term can be further derived into  $$\log(P(y_i|x_i,z^i_s))= c+\frac{||y_i-f(x_i,z_i^s) ||^2_2}{\sigma^2}$$ where $c$ is a constant term independent of $x_i,y_i,z^i_s$

\begin{figure*}[h]
	\centering
	\includegraphics[width=1\linewidth]{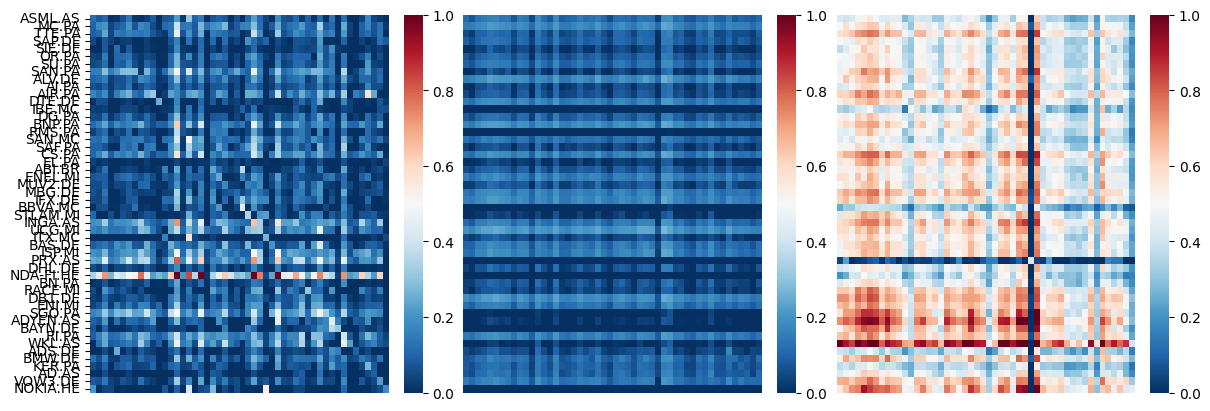}
	\caption{Cross correlation from data (left), ACP estimate of cross correlation (centre), and CAP estimate of cross correlation (right)}
	\label{fig:corr2}
\end{figure*}
\subsubsection{Architecture and optimisation techniques}
As for the architecture of the neural networks (formally speaking the space $F_1, F_2$), we use two layers of RELU network for the encoder function, with dimensions linking the input space ($\mR^p \times \mR^d$) to $\mR^q \times \mR^q$ for the first layer, an untrained identity map from $\mR^q \to \mR^q$ to map to the expectation term and an untrained softplus from $\mR^q \to (0,\infty)^q$ to map the variance.

We use two layers of RELU and one layer of linear network for the decoder function. The dimensionality depends on the dimensionality of the input --- when $d=50$ (the MCVAE model), the dimensions for the two layers of RELU are input dimension to 64 ($\mR^{59}\times \mR^{64}$), followed by $\mR^{64} \times \mR^{64}$ and then linear layer  $\mR^{64} \times \mR^{50}$ for output. When $d=1$ (the UCVAE model), the layers are $\mR^{27}\times \mR^{16}$ and $\mR^{16}\times \mR^{8}$ for RELU, then a linear layer of $\mR^{8}\times \mR$ for output.

The optimisation procedure can be summarised as a combination of ADAM (default setting in tensorflow version 2.16.1) and validation early stopping. The training is hated when the validation loss exceeds 1\% of the local minimum within the last 3 steps of the training.

\subsubsection{Technical note regarding $\sigma$ calibration}
There are various ways to estimate or calibrate $\sigma$. As the focus of this paper is more on the point forecast $y_{t+k|t}$ instead of interval forecasting, we aim to explore the centre of sampling as opposed to the tails, hence calibration is used for the value of $\sigma$. In section \ref{SoP}, we used $\sigma=0.1$ for efficient sampling to produce overall results, and in section \ref{DaG} and \ref{Discuss}, we used $\sigma=1$ to have wider range of samples and to further the investigation to correlation estimates.

\bibliographystyle{ACM-Reference-Format}
\bibliography{bibliography}
\end{document}